\documentclass[prl,twocolumn] {revtex4}
\usepackage{graphicx}
\usepackage{amsfonts}
\usepackage{amssymb}

\newcommand{\ins}{{\mbox{\tiny{in}}}}
\newcommand{\out}{{\mbox{\tiny{out}}}}
\newcommand{\is}{{\mbox{\tiny{is}}}}

\begin{document}
\author{Eytan Grosfeld,$^{1}$ Steven H. Simon,$^{2}$ and Ady Stern$^{1}$}
\affiliation{$^1$Department of Condensed Matter, Weizmann
Institute of Science, Rehovot 76100, Israel \\ $^2$Bell
Laboratories, Lucent Technologies, 600 Mountain Avenue, Murray
Hill, New Jersey 07974}
\title{Switching noise as a probe of statistics in the fractional quantum Hall effect}
\begin{abstract}

We propose an experiment to probe the unconventional quantum
statistics of quasi-particles in fractional quantum Hall states by
measurement of current noise. The geometry we consider is that of
a Hall bar where two quantum point contacts introduce two
interfering amplitudes for back-scattering. Thermal fluctuations
of the number of quasi-particles enclosed between the two point
contacts introduce current noise, which reflects the statistics of
the quasi-particles. We analyze abelian $\nu=1/q$ states and the
non-abelian $\nu=5/2$ state.

\end{abstract}
\maketitle

The fractional quantum Hall effect (FQHE) is characterized by
unconventional properties of its quasi-particles. In particular,
the quasi-particles carry a charge which is a fraction of the
electron's charge and satisfy a fractional statistics. In abelian
quantum Hall states, such as the $\nu=1/3$ state, the winding of
one quasi-particle around another is associated with a phase shift
to the wave function which is a fraction of $2\pi$. Even more
interesting are non-abelian states, such as the $\nu=5/2$ state:
when quasi-particles are present in the ground state, due to a
slight deviation of the filling factor from $\nu=5/2$, the ground
state becomes degenerate. Then, a winding of one quasi-particle
around another is associated with a unitary transformation in the
subspace of ground states, leading to non-abelian statistics.

Probing the statistics  is difficult in both the abelian and the
non-abelian cases, as one should devise experiments that are
sensitive to the non-local statistical interaction between two
such quasi-particles. Recent theoretical works suggest that
interference experiments in a two point contact geometry may serve
as probes of statistics. Such experiments were suggested in the
context of the abelian FQHE \cite{bib20}, the non-abelian
$\nu=5/2$ \cite{bib21,bib22,bib23}, and the non-abelian $\nu=12/5$
\cite{bib201,bib202}.

\begin{figure}[h]
    \centering
    \includegraphics[width=120pt,angle=270]{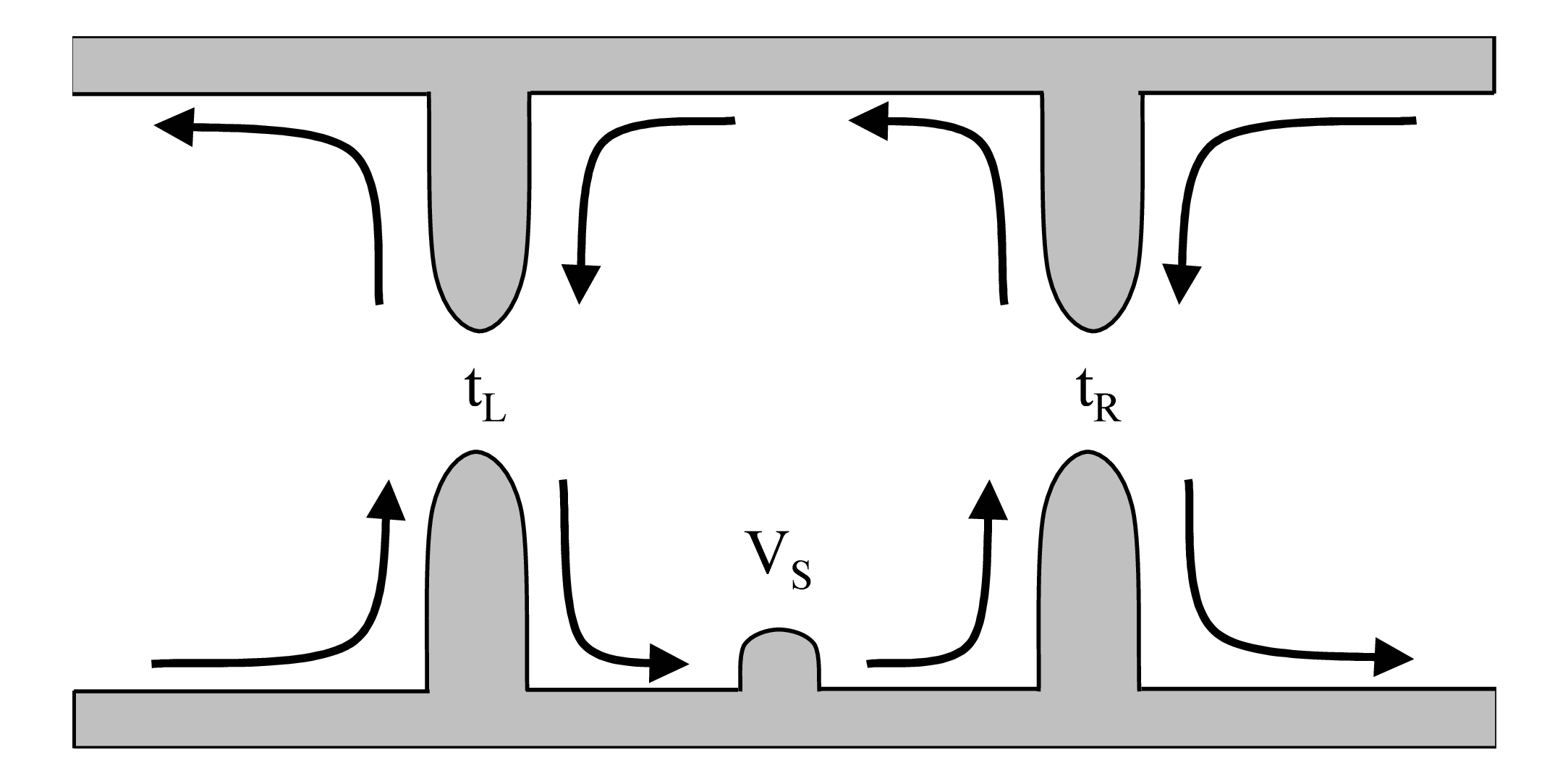}
     \caption{Experimental setup shared by recently proposed interference experiments.
     Quasi-particles arrive from the left, along the lower edge,
     and back-scatter from the left or right quantum point
     contacts with tunneling amplitudes $t_L$ and $t_R$. The area of the island can be controlled by varying the voltage on a side gate.}
    \label{fig:noise}
\end{figure}

The geometry considered in these experiments (see Fig.
[\ref{fig:noise}]) is a Hall bar with two quantum point contacts.
We shall refer to the area included between these two quantum
point contacts as the "island". The left and right quantum point
contacts introduce weak back-scattering of quasi-particles between
the two edges, characterized by tunneling amplitudes $t_L$ and
$t_R$. The back-scattered current, driven by a source-drain
potential $V_{sd}$, is measured. Its magnitude is determined by
the interference of the two tunneling amplitudes. An oscillating
interference pattern may be observed when the area of the island
is varied by means of a side gate.

The fractional statistics of the FQHE quasi-particles manifests
itself in the dependence of the interference pattern on the number
of quasi-particles localized within the island, denoted by $n_{\rm
is}$\cite{bib20,bib21,bib22,bib23,bib201,bib202}. This number may
be varied either by means of an anti-dot within the center of the
island, or by a variation of the magnetic field.  For the abelian
states ($\nu=1/q$) the quasi-particles introduce a phase shift: a
change of $n_{is}$ by one leads to a phase shift of $2\pi/q$ in
the interference pattern\cite{bib20}. More interestingly, for the
non-abelian $\nu=5/2$  the interference term is turned off when
$n_{\rm is}$ is odd, and is turned on when $n_{\rm is}$ is
even\cite{bib22,bib23}. The phase of the interference term when
$n_{\rm is}$ is even is a subtle issue, to be elaborated on below.
At zero temperature, and when the bulk chemical potential is
within a range of localized states, the number $n_{\rm is}$ is
determined by disorder, interactions and magnetic field, and is
time independent. Thermal fluctuations, however, may make that
number time dependent.

In this paper we consider the effect of thermal fluctuations of
$n_{\rm is}$. We show that if the characteristic time scale of the
fluctuations is long enough (though much shorter than the time of
the experiment), they do {\it not} wash out effects of abelian and
non-abelian statistics. We show that in this case the
back-scattered current becomes time dependent, due to the time
dependence of $n_{\rm is}$. This time dependence is a source of a
unique type of current noise, whose properties reflect the
fractional statistics of the quasi-particles. We analyze the
conditions under which this source of noise is larger than shot
noise, and explain how it may be distinguished from other noise
sources.

For a given $n_{\is}$ the conductance for the abelian states is
given by \cite{bib20}
\begin{eqnarray}
    \label{eq:abelian} G=G_0[1+\beta\cos(\phi_0+2\pi n_{\is}/q)]
\end{eqnarray}
and for the $\nu=5/2$ state it is given by \cite{bib22,bib23}
\begin{eqnarray}
    \label{eq:non-abelian} G=\left\{\begin{array}{ll}
    G_0 & (\mbox{$n_{\is}$ odd})\\
    G_0[1+\beta\cos(\phi_0+\pi n_{\is}/4+r \pi)] & (\mbox{$n_{\is}$ even})\end{array}\right.
\end{eqnarray}
Here $G_0$ is the conductance in the absence of interference
$G_0\propto |t_L|^2+|t_R|^2$, and $\beta\propto |t_Rt_L|/G_0$ is
the visibility of the interference, both calculated to lowest
order in the tunneling amplitudes. The parameter $r$ assumes the
value $0$ or $1$; the probability for the two values is equal, and
as long as $n_{\is}$ does not vary, $r$ remains fixed as well. Its
behavior when $n_{\is}$ varies is studied below. The phase
$\phi_0$ is independent of $n_{\is}$ and may be controlled by
changing the area of the island by means of a side gate. At finite
temperatures, the conductance $G$ becomes time-dependent due to
hopping of quasi-particles through the quantum point contact,
leading to fluctuations in $n_{\is}$. Measurement of the
conductance would reveal these fluctuations. If the hopping events
are slow compared with the measurement time, one could observe
single hopping events, manifested as abrupt changes of the
conductance between the allowed values prescribed by Eqs.
(\ref{eq:abelian}) and (\ref{eq:non-abelian}) giving clear
evidence of the statistics. However, if the measurement time is
larger than the hopping time, the measured conductance would
approximately average Eqs.
(\ref{eq:abelian}),(\ref{eq:non-abelian}), while the fluctuations
in $n_{\is}$ would result in current noise. We now proceed to
analyze this noise, and show that valuable information about the
statistics of the quasi-particles remains encoded into that noise.

We consider two scenarios for the fluctuations of $n_{\is}$. In
the first, a single quasi-particle hops between two trapping
centers located on the two sides of the tunneling point of one of
the quantum point contacts. Consequently, $n_{\is}$ fluctuates
between two consecutive integers, $n_0$ and $n_0+1$. For abelian
states, the conductance fluctuates between two values, and the
noise reflects these fluctuations. For non-abelian states, $n_{\rm
is}$ fluctuates between an even and an odd number, and the
interference term is turned on and off by the fluctuations.
Remarkably, in this case we find the noise to depend on the parity
of $n_0$.

The second scenario is that in which the fluctuations in $n_{\rm
is}$ are much larger than one. Surprisingly, even this limit holds
interesting consequences for the measured noise, both for
$\nu=1/q$ and $\nu=5/2$.

In all the cases we consider the system has several possible
values for the conductance. Quasi-particle hopping leads to random
and abrupt switches of the conductance between the allowed values,
that result in telegraph-type noise in the measured conductance
$G(t)$ (note that we define the conductance here as the ratio of the {\it back-reflected} current to $V_{ds}$). We assume certain rules and transition rates for the
hopping of quasi-particles, and compute the current noise by
calculating the two-time correlation function $S_2(t)=\langle
\tilde{G}(t)\tilde{G}(0)\rangle V_{sd}^2$ and the three-time
correlation function $S_3(t,t')=\langle \tilde{
G}(t')\tilde{G}(t)\tilde{G}(0)\rangle V_{sd}^3$ where
$\tilde{G}(t)=G(t)-\langle G\rangle$ and $\langle G\rangle$ is the
average conductance. We transform to frequency space using
\begin{eqnarray}
    \label{eq:correlator-2}
    S_2(\omega)&=&\int dt e^{i \omega t} S_2(t) \\
    \label{eq:correlator-3}
    S_3(\omega,\omega')&=&\int d t d t' e^{i \omega (t'-t)}e^{i \omega'(t+t')/2}S_3(t,t')
\end{eqnarray}
We first disregard the contribution of other sources of noise to
the correlators. Those are discussed towards the end of the paper.

We calculate these current correlators by standard methods for
Markov chains. The states of the model are labelled $1,2,\ldots$
with conductances $G_1,G_2,\ldots$ and corresponding probabilities
$P_1,P_2,\ldots$. We characterize the process by a diagram whose
vertices are the states while the edges, which are directed and
weighted, describe allowed transitions with corresponding rates.
This diagram is encoded in the transition rate matrix $M$, whose
element $ M_{ij}$ describes the rate of transitions between states
$i$ and $j$. The $M$-matrix has two defining properties: It is
conservative $M_{ii}=-\sum_{j\neq i} M_{ij}$, and
$\mathbf{P}_0=(P_1,P_2,\ldots)^{\mbox{\small{tr}}}$ satisfies $M
\mathbf{P}_0=0$. Given that we start in state $j$ at time $t=0$,
the probability $P_{ij}(t)=P(G(t)=G_i|G(0)=G_j)$ that we end in
state $i$ at time $t$, is given by the matrix equation
\begin{eqnarray}
    \partial_t P=M P
\end{eqnarray}
Its solution is $P_{ij}(t)=\langle i|e^{M t}|j\rangle$,
where $|i\rangle$ is a unit vector whose $i$'th entry is $1$. The conditional probabilities $P_{ij}$ satisfy $P_{ij}(0)=\delta_{ij}$ and, for all $j$, $P_{ij}(t\to\infty)=P_i$. Consequently, for $t'>t>0$, the correlators $S_2$
and $S_3$ may be written as
\begin{eqnarray}
    S_2(t)&=&V_{sd}^2\sum_{ij}\tilde{G}_i \langle i|e^{M t}|j\rangle \langle j|\mathbf{P}_0\rangle \tilde{G}_j  \\
    \nonumber S_3(t,t')&=&V_{sd}^3\sum_{ijl} \tilde{G}_i \langle i|e^{M (t'-t)}|j \rangle \tilde{G}_j \langle j|e^{M
    t}|l\rangle \langle l|\mathbf{P}_0\rangle \tilde{G}_l
\end{eqnarray}
where $\langle i|\mathbf{P}_0\rangle=P_i$, and $\tilde{G}_i=G_i-\sum_i P_i
G_i$ is the conductance measured with respect to the average
conductance.

\begin{figure}[h]
    \centering
    \includegraphics[width=180pt,angle=270]{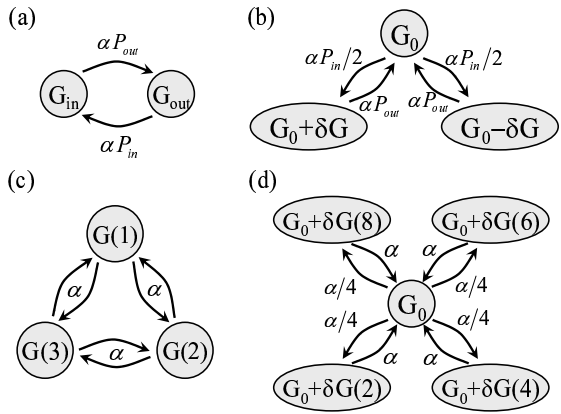}
     \caption{The state space of the Markov chains considered in the text. (a) The abelian FQHE with one hopping quasi-particle, and the non-abelian state with $n_0$ even.
     (b) The non-abelian state with $n_0$ odd. (c) The abelian $\nu=1/3$ with a
     flow of quasi-particles in and out of the island. (d) The
     non-abelian case with a flow of quasi-particles.}
    \label{fig:diagrams}
\end{figure}

For the case when thermal fluctuations make a single
quasi-particle hop in and out of the island near the point
contact, we assume hopping events to be uncorrelated, with hopping
rates $\alpha P_\ins$ and $\alpha P_\out$ into and out of the
island respectively. The probability for the quasi-particle to be
inside the island is $P_\ins$, and for it to be outside the island
is $P_\out=1-P_\ins$.

We start with the $\nu=1/q$ state, for which the two conductance
values are $G_\out$ with probability $P_\out$ and $G_\ins$ with
probability $P_\ins$. The transition rates for this case are
depicted in Fig. [\ref{fig:diagrams}a]. The correlation functions
are given by
\begin{eqnarray}
    \label{eq:noise-2-1}
    S^a_2(\omega)&=&\frac{2 \alpha P_\ins(1-P_\ins)\delta G^2V_{sd}^2}{\alpha^2+\omega^2}\\
    \label{eq:noise-2-2}
    S^a_3(\omega,\omega')&=&P_\ins(1-P_\ins)(1-2P_\ins)f(\omega,\omega') \delta G^3 V_{sd}^3
\end{eqnarray}
where $\delta G=G_\ins-G_\out$, and
\begin{eqnarray}
    \nonumber f\left(\omega,\omega'\right)=\frac{\alpha^2(4\omega^2+3{\omega'}^2+12\alpha^2)}{2(\omega'^2+\alpha^2)((\omega+\frac{\omega'}{2})^2+\alpha^2)((\omega-\frac{\omega'}{2})^2+\alpha^2)}
\end{eqnarray}
 When
compared to shot noise, where $S_2(\omega=0)=2 e^*(P_\ins
G_\ins+P_\out G_\out)V_{sd}$ and $S_3=0$, we see that if
$P_\ins\approx 1/2$ and the visibility $\beta$ is close to unity,
the contribution (\ref{eq:noise-2-1}) typically becomes larger
than the shot noise once the rate $\alpha$ is smaller than the
rate at which quasi-particles cross the device, $G_0 V/e^*$.

For $\nu=5/2$ we find a surprising dependence of the noise on the
parity of $n_0$. This dependence originates from the subtle way in
which the phase of the interference term is determined when
$n_{\rm is}$ is even. In that case there are two possible
interference patterns, mutually shifted by a phase of $\pi$,
corresponding to the two possible values of $r$ in Eq.
(\ref{eq:non-abelian}).

We now analyze the way this parameter varies when $n_{\rm is}$
fluctuates by the hopping of one quasi-particle. We show that when
$n_0$ is even, $r$ remains fixed for the duration of the
experiment, such that the fluctuations of $n_{\rm is}$ translates
to a conductance that switches between two possible values, with
transition rates depicted in Fig. [\ref{fig:diagrams}a]. Then, the
second and third cumulants of the conductance are  given by Eqs.
(\ref{eq:noise-2-1}) and (\ref{eq:noise-2-2}) with $\delta
G=G_0\beta\cos\phi$.

In contrast, when $n_0$ is odd the hopping of the quasi-particle
randomizes $r$, and thus translates to a switching of the
conductance between {\it three} possible values: one value, with a
probability $P_\ins$, is $G_0$. The other two values, with equal
probabilities $P_\out/2$, are $G_0[1\pm\beta\cos(\phi)]$. The
transition rates are depicted in Fig. [\ref{fig:diagrams}b],
yielding
\begin{eqnarray}
    \label{eq:noise-3-1}
    && S^{b}_2(\omega)=\frac{2 \alpha P_\ins\delta
    G^2}{(1-P_\ins)^2\alpha^2+\omega^2}\\
    \label{eq:noise-3-2}
    && S^{b}_3(\omega,\omega')=0
\end{eqnarray}

The dependence of the conductance on the parameter $r$ reflects
the degeneracy of the ground state in the presence of the $n_{\rm
is}$ localized quasi-particles, a degeneracy which is the
cornerstone of non-abelian statistics. We now review that
dependence shortly\cite{bib21,bib22,bib23}, before analyzing the
way it is affected by fluctuations in $n_{\rm is}$.

The $\nu=5/2$ state may be described as a $p$-wave superconductor
of composite fermions \cite{bib2}, with the quasi-particles being
vortices in that super-conductor. Each vortex carries a localized
zero energy core state, which is a Majorana fermion $\gamma$,
satisfying $\gamma^\dag=\gamma$, and $\gamma^2=1$. The zero energy
core-states make the ground state degenerate. Mutual winding of
vortices induces a transformation within the subspace of
degenerate ground states. As explained in \cite{bib22}, when a
quasi-particle (denoted by a subscript $edge$) that moves along
the bottom edge of the system in Fig. [\ref{fig:noise}], from
$x=-\infty$, is back-reflected by the two quantum point contacts,
the interference term of the left and right back-reflections is
multiplied by the expectation value $\langle {\rm
g.s.}|\Gamma|{\rm g.s. }\rangle$, with
\begin{equation}
\Gamma\equiv\prod_{i=1}^{n_{\rm is}}(\gamma_{edge}\gamma_i)
\label{gammadef}
\end{equation}
where the $\gamma_i$'s are the Majorana operators associated with the quasi-particles localized on the island, and $\gamma_{edge}$ is the Majorana operator of the back-reflected quasi-particle.
When $n_{\rm is}$ is even, $\Gamma$ depends only on the $n_{\rm is}$ Majorana states localized in the island, and the expectation value is the same for all back-reflected quasi-particles. Then, the measurement of the
conductance collapses the system into an eigenstate of
$\Gamma$ \cite{bib3}. This operator has two
eigenvalues and each one of them corresponds
to a different value of $r$. When $n_{\rm is}$ is odd, $\Gamma$ is different for different back-scattered quasi-particles, and the interference term averages to zero.

Let us now carefully examine how the value of $r$ is affected by
the hopping of one quasi-particle into and out of the island.
Consider first a system that fluctuates between an {\it even}
$n_0$ and an {\it odd} $n_0+1$ quasi-particles in the island. When
there are $n_0$ quasi-particles and the conductance is measured,
the system collapses into an eigenstate of $\Gamma$. When the
$n_0+1$'th quasi-particle hops in from outside the island, each
following event of back-reflection would apply unitary
transformations on $|{\rm g.s.}\rangle$, of the form
$\gamma_{n_{0}+1}\gamma_{edge}\Gamma$. Since these transformations
commute with $\Gamma$, the resulting state would still be an
eigenstate of $\Gamma$, with the same eigenvalue. Thus, when the
${n_{0}+1} $'th quasi-particle hops out of the island again, the
value of the parameter $r$ is left unchanged. The hopping of the
quasi-particle into and out of the island switches the system
between two values of the conductance, leading to a two-states
Markov process.

The situation is different when $n_0$ is odd, and $n_0+1$ is even. Again, following the hopping of the $n_{0}+1$'th quasi-particle outside the island, unitary transformations of the form $\gamma_{n_{0}+1}\gamma_{edge}\Gamma$ are applied on the ground state, but this time, since $\gamma_{n_0+1}$ appears in $\Gamma$, we have,
\begin{equation}
[\Gamma, \gamma_{n_{0}+1}\gamma_{edge}\Gamma]\ne 0
\label{commrel}
\end{equation}
Thus, in this case the back-reflection changes the state $|{\rm g.s.}\rangle$ into a state that is not an eigenstate of $\Gamma$, and randomizes the value of $r$. Consequently, the hopping of one quasi-particle switches the conductance between three values: when the quasi-particle is outside of the island, the conductance is $G_0$. When it is inside the island, the conductance assumes the values $G_0[1\pm\beta\cos(\phi)]$, with equal probability.

We now turn to the second case where the fluctuations in $n_{\rm
is}$ are much larger than one. We start with $\nu=1/q$. The
conductance assumes one of the values $G(m)=G_0[1+\beta
\cos(\phi_0+2\pi m/q)]$, where $m=0,1,\ldots,q-1$ is $n_{\rm is}$
modulo $q$. The system shifts between two adjacent $m$ values at
equal rates. Consequently, the $M$-matrix of this problem is that
of a tight-binding model with $q$ sites and periodic boundary
conditions; its eigenvalues are $-(1-\cos(k))$, where
$k=0,2\pi/q,\ldots,(q-1)2\pi/q$. We can therefore Fourier
transform and write the correlation function as
\begin{eqnarray}
    S_2^c(t)=\sum_{xy,k}\tilde{G}_x\tilde{G}_y e^{i
    k(y-x)}e^{-\alpha t (1-\cos(k))}
\end{eqnarray}
Consequently, we get
\begin{eqnarray}
    S^c_2(\omega)=\frac{\alpha (\beta G_0)^2}{\alpha^2\left[1-\cos\left(2\pi/q\right)\right]^2+\omega^2}
\end{eqnarray}
and $S^c_3$ is zero.

For $\nu=5/2$ we get the following picture. When the instantaneous
number of quasi-particles in the island is odd, no interference
pattern occurs, and the conductance is $G_0$. When it is even, the
state of the system collapses into one of the eigenvalues of
$\Gamma$ with equal probability. The possible values of the
conductance for even $n_{\rm is}$ are $G_0[1\pm
\beta\cos(\phi_0)]$, $G_0[1\pm \beta \sin(\phi_0)]$, with equal
probabilities. The diagram describing the Markov process consists
of the odd particle number state in the middle and four branches
going out to each of the even states. The resulting two-time
correlation function is
\begin{eqnarray}
    S^d_2(\omega)=\frac{1}{2}\frac{\alpha(\beta G_0)^2}{\alpha^2+\omega^2}
\end{eqnarray}
Three-time correlations again vanish.

Finally, we turn to discuss other mechanisms of current noise and possible methods
to differentiate between them and the noise we focus on here.

At high frequencies, the noise spectrum is typically dominated by
shot noise, given by $2 e^* \langle I_B\rangle$ where $\langle
I_B\rangle$ is the average back-reflected current. The linear dependence of shot noise on current is in sharp contrast to the $\langle I_B \rangle^2$ dependence of the noise we discuss here. Furthermore, the noise we discuss here is a consequence of interference of two amplitudes for back-reflection. Thus, it is  turned off when only one point contact is active, in contrast to shot noise, which is present also for a single point contact.

At low frequencies, the noise spectrum will be dominated by $1/f$
noise due to random motion of impurities. This type of noise may
show similar characteristics to the noise we discuss here. Its
existence is usually attributed to resistance changes due to
activated processes with uniformly distributed activation energies
over a small range \cite{bib4}. However, while $1/f$ noise is common to integer and fractional quantum Hall states, the contribution we discuss here is unique to the FQHE states. if one tunes the system to an integer
filling factor, say $\nu=2$, the contribution to the noise
described in this paper should disappear. All other sources of
noise (save the shot noise) remain intact, and this method can be
used to isolate the statistical noise unique to the fractional
states.

To conclude, we showed that charge fluctuations on the island
result in a novel type of noise unique to the fractional quantum
Hall effect. For the non-abelian case, we found an even-odd effect
for the noise spectrum which has its origins in the non-abelian
properties of the quasi-particles. Consequently, the noise
spectrum of the back-scattered current may serve as a probe of the
unique properties of the abelian FQHE states and the non-abelian $\nu=5/2$ state.

We acknowledge financial support from the US-Israel BSF (2002-238)
and the Minerva foundation.

\end{document}